# Towards Bose-Einstein condensation of excitons in an asymmetric multi-quantum state magnetic lattice

A. Abdelrahman[1*], M. Vasiliev[1], K. Alameh[1], P. Hannaford[2], Byoung S. Ham[3], and Yong-Tak Lee[4]

1 *Electron Science Research Institute, Edith Cowan University, 270 Joondalup Drive, Joondalup WA 6027 Australia.*
2 *Centre for Atom Optics and Ultrafast Spectroscopy, and ARC Centre of Excellence for Quantum Atom Optics,*
*Swinburne University of Technology, Melbourne, Australia 3122*
3 *Center for photon information processing, and the Graduate school of information and telecommunications,*
*Inha University, Incheon 402-751, S. Korea*
4 *Department of Information and Communications, Gwang-ju Institute of Science and Technology, Gwang-ju, 500-712, S.Korea*
* aabdelra@student.ecu.edu.au

**Abstract**
An asymmetric multi-quantum state magnetic lattice is proposed to host excitons formed in a quantum degenerate gas of ultracold fermionic atoms to simulate Bose-Einstein condensation (BEC) of excitons. A Quasi-two dimensional degenerate gas of excitons can be collected in the in-plane asymmetric magnetic bands created at the surface of the proposed magnetic lattice, where the ultracold fermions simulate separately direct and indirect confined electron-hole pairs (spin up fermions-spin down fermions) rising to the statistically degenerate Bose gas and eventually through controlled tunnelling to BEC of excitons. The confinement of the coupled magnetic quantum well (CMQWs) system may significantly improve the condition for long lived exciton BEC. The exciton BEC, formed in CMQWs can be regarded as a suitable host for the multi-qubits (multipartite) systems to be used in quantum information processors.

## I. INTRODUCTION

In a semiconductor or an insulator, an exciton is a quasi-particle formed as a bound state of an electron-hole pair with an overall zero charge. Such quasi-particles behave as bosons, particles with integer spin, and obey Bose-Einstein statistics. The composite Bose quasi-particle (excitons) in the low density limit become similar to a hydrogen atom; therefore by reducing the temperature or increasing the exciton density, the occupation of the low energy state will increase leading to a quantum degenerate gas that condenses into a single quantum state (BEC) known as an exciton Bose-Einstein condensate [1]. In a similar way to an atomic Bose-Einstein condensate, an exciton BEC can exhibit macroscopic quantum mechanical phenomena such as superfluidity and quantum interference. The mass of the excitons formed in a semiconductor is small compared to the free electron mass. Thus, a quantum degenerate gas of exciton BEC occurs at a temperature around 1 K, higher by many orders of magnitude compared to the temperature of a Bose-Einstein condensation made of ultracold atoms. This attracted considerable attention due to its ability to produce an exciton BEC in semiconductor devices [2]. However, the life time of an exciton BEC, formed using low dimensional semiconductors, is very short, of the order of nanoseconds, before electron-hole recombination occurs. Thus, it is extremely difficult to reach the long-lived exciton BEC degeneracy. Moreover, it is very hard to reach the required temperature in practice, because the exciton temperature exceeds the temperature of the semiconductor lattice [3].

## II. ASYMMETRIC MULTI-QUANTUM STATE MAGNETIC LATTICE

Recently, several micrometer-scale structures generating periodic magnetic potentials have been proposed for trapping cold atoms as an alternative to optical lattices. These specifically engineered quantum devices can be realized by manufacturing one dimensional and two dimensional periodic structures on a device called *Atom Chip* [4,5]. The proposed structure is shown in Figure 1(a), where the spatial magnetic field components $B_x$, $B_y$ and $B_z$ can be written analytically as a combination of a field decaying away from the surface of the trap in the $z$ direction and the periodic magnetic field distribution in the $x$-$y$ plane produced film. The unpatterned area (*magnetic wall*) surrounding the $n \times n$-hole matrix is the key element for controlling the energy levels and the magnetic field minima of the trap [4]. The presence of holes results in a magnetic field distribution whose non-zero local minima are located at effective z-distances from the holes.

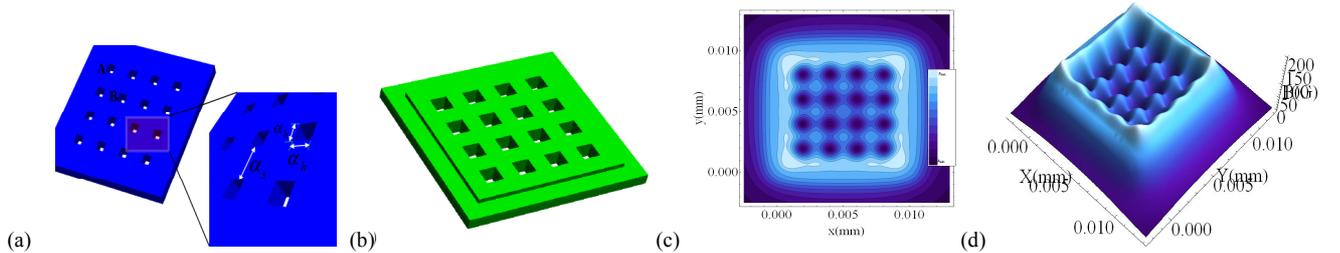

**Fig. 1.** (a) (b) Proposed nano-structured garnet-Based magnetic lattice for confinement and trapping of ultra-cold atoms. It consists of an $m \times m$ array of $n \times n$-hole blocks, where each block is surrounded by contiguous film surface areas that create magnetic "walls" ($m = 2$ and $n = 4$ in the shown structure). $α_h$ represents the hole dimension while $α_s$ represents the period between two holes. The sites **A** and **B** represent two magnetic quantum wells in two different magnetic bands. (c) (d) Contour and 3D plot of a localized magnetic field distribution across the $4 \times 4$ magnetic lattice.

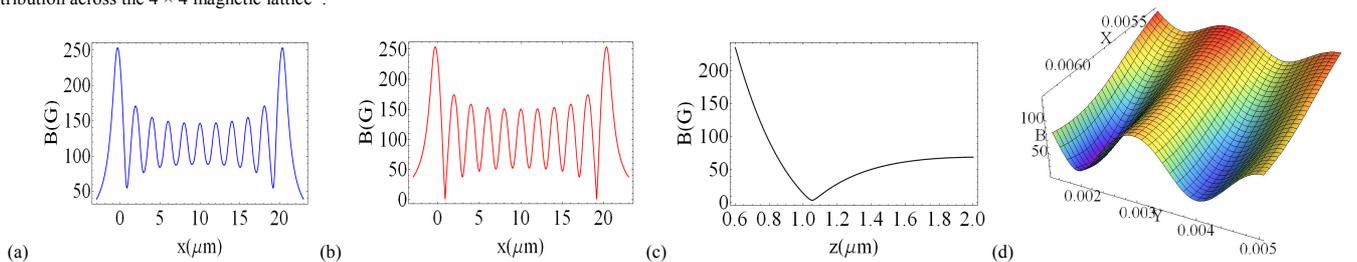

**Fig. 2.** Magnetic field distribution across $10 \times 10$ magnetic lattice (a) at the center and (b) at the edge of the trap along x-axis. (c) The non-zero magnetic minima at effective distance from the trap surface (c) 3D plot of two indirectly coupled quantum wells. The magnetic barrier between the vertical sites represent the magnetic band gap which can suppressed using external magnetic field $B_{z-bias}$.

These minima are localized in small volumes representing the potential wells that contain a certain number of quantized energy levels for the cold atoms to occupy. In our design, we set the size of the holes $\alpha_h$ and the separation of the holes $\alpha_s$ as $\alpha_h = \alpha_s \equiv \alpha$ to simplify the mathematical derivation. The magnitude of the total magnetic field at the effective distance of the localized potential wells can be described by including the effect of the magnetic field components of the bias fields along the $x$, $y$ and $z$ directions, $B_{x\text{-}bias}$, $B_{y\text{-}bias}$ and $B_{z\text{-}bias}$. However we use only $B_{z\text{-}bias}$ to produce the CMQWs. For an infinite magnetic lattice the magnetic magnitude $B$ is described as [4]

$$B = \left\{ B_{x\text{-}bias}^2 + B_{y\text{-}bias}^2 + B_{z\text{-}bias}^2 + 2B_o^2\left(1 - e^{-\beta\tau}\right)^2 e^{-2\beta[z-\tau]}\left[\cos(\beta x)\cos(\beta y)\right] + 2B_o^2\left(1 - e^{-\beta\tau}\right)e^{-\beta[z-\tau]} \times \left(\left[B_{x\text{-}bias} + B_{z\text{-}bias}\right]\cos(\beta x) + \left[B_{y\text{-}bias} + B_{z\text{-}bias}\right]\cos(\beta y)\right) \right\}^{1/2} \quad (1)$$

where $B_o = \mu_o M_z/\pi$ is the magnetic induction at the surface of the film, $\tau$ is the film thickness and a plane of symmetry is assumed at $z=0$, $\beta = \pi/\alpha$. By configuring the magnetic walls around the lattice as shown in Figure 2, we can control the curvature direction of the CMQWs to be outward on inward as shown in Figure 3(a). The resulting magnetic lattice sites are spatially distributed in the x-y plane with n-bands of non-zero local magnetic minima. Each magnetic band contains uni-directional *adjacent* and symmetrical coupled magnetic quantum wells along the x-axis and y-axis, while vertically along the x-z and y-z directions they create indirectly coupled quantum wells separated by magnetic barriers defined as *magnetic band gaps*. The number and the values of the magnetic band gaps are determined by the number of holes distributed periodically in the x-y plane with their plane of symmetry at z = 0. In the simulated case shown in Figure 3, the gap between each band increases outward from the center of the magnetic lattice as shown in Figure 3(b) due to the magnetic wall effect. The gap value and the magnetic barrier between the sites $\Delta B$ can be controlled via the application of an external magnetic bias field along the negative z-direction. This causes a reduction in the effective thinfilm magnetization $M_z$, and consequently affects the magnetic band coupling parameters.

## III. FORMATION OF EXCITONS BEC IN MAGNETICALLY TRAPPED ULTRACOLD FERMI GASES

The energy band diagram of coupled quantum wells in a semiconductor is shown in Figure 4(a). The indirectly coupled electron-hole pair forms an indirect exciton. In the proposed asymmetric multi-quantum states magnetic lattice, the magnetic quantum wells, which are the horizontal lattice sites within each magnetic band, are expected to simulate the adjacent multi-quantum well system in the valence band and the conduction band of the semiconductor separated by the energy band gap which in this case is represented by the magnetic band gap. The *indirectly coupled* magnetic quantum wells are assumed to be two lattice sites vertically distributed along two different magnetic bands and separated by the magnetic band gap. These vertical quantum wells can be coupled by applying an external magnetic field, $B_{z\text{-}bias}$, along the negative z-axis.

The ultracold fermions are distributed along the n different magnetic bands by occupying the available energy level of each magnetic quantum well (lattice site). The overall *surface-spin* at each magnetic band will screen the fermion atoms' spin at the final energy level of each magnetic site. In the case of uncoupled vertical magnetic quantum wells, the adjacent magnetic lattice sites are assumed to be spin independent from the nearest neighbour-sites in each magnetic band, as well as from those sites in the other magnetic bands. By applying a $B_{z\text{-}bias}$ the magnetic barrier between adjacent sites can be lowered while at the same time minimizing the magnetic band gaps. Tunnelling between adjacent magnetic sites is not allowed due to the availability of the energy level at each site and hence reaching a spin-decoupled state. The temperature will be reduced at each magnetic band and the hot fermions will tunnel to the lowest non-zero magnetic minima available in the bottom bands. At a certain $B_{z\text{-}bias}$ value, the quantum wells at the bottom magnetic band will reach their capacity, where the application of $B_{z\text{-}bias}$ will not increase the non-zero local minima at any magnetic band. Bose-Einstein condensation of the excitons is expected to occur at a certain value of $B_{z\text{-}bias}$. Starting from the top magnetic band, the surface-spin of the fermions occupying the final energy level of each magnetic quantum well will be coupled, following the Pauli principle, to the surface-spin of the fermions of the bottom magnetic band. The surface-spin of the fermions in the bottom magnetic band are also occupy the final energy state of the magnetic quantum well of the bottom band. As a result, direct-excitons are created between each two magnetic bands, and due to the reduction of the temperature in the upper magnetic bands and the increase of the fermions density in the bottom magnetic bands, the occupation numbers of the low-energy states will increase leading to degenerate quantum gases of excitons and eventually to Bose-Einstein condensation of excitons. The exciton BEC is expected to be in a long-lived bound state due to the pronounced reduction of temperature on the upper magnetic bands (conduction) and the enhanced density of fermions in the bottom magnetic band (valence). It is also possible to create indirect-excitons by applying $B_{x\text{-}bias}$ or $B_{y\text{-}bias}$.

## IV. CONCLUSION

Current research in this area is concentrated on the long-lived exciton BEC using semiconductors, where the exciton lifetime should exceed the recombination rate of the electron-hole pairs which is characterized by the cooling time of the thermal excitons that are optically generated in semiconductor device. However this is experimentally very difficult and the temperature of the excitons exceeds the semiconductor lattice temperature. Our proposed magnetic structure is expected to offer long-lived exciton BEC as described above. The coherence time is also expected to be long making the approach very suitable to maintain a host for multi-qubits which is recommended to be used for multipartite entanglement in the field of quantum information processing using magnetically trapped ultracold atoms.

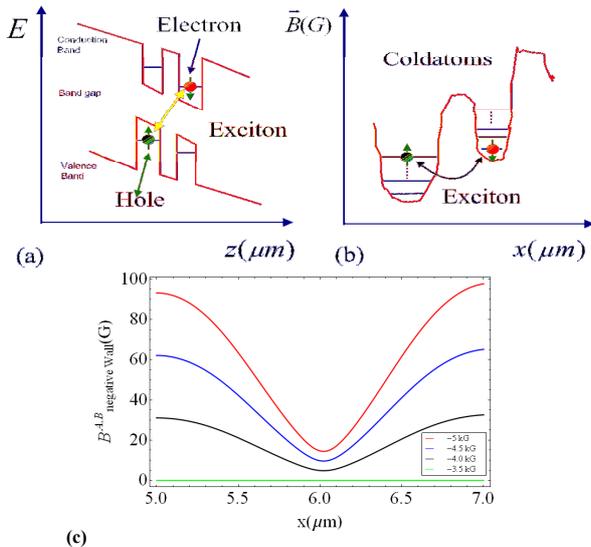

**Fig. 3.** (a) Energy band diagram of coupled quantum wells in a semiconductor device with indirect-exciton formed by electron-hole bound state and (b) Schematic diagram of two magnetic sites of the asymmetric multi-quantum state magnetic lattice which are expected to simulate both direct and indirect-excitons by induced spin coupling of ultracold fermions in two different magnetic bands. (c) Magnetic barrier between two sites can be minimized by applying external magnetic field $B_{z\text{-}bias}$.